\documentclass[runningheads]{llncs}
\usepackage{tikz}
\usepackage{graphicx}
\usepackage{amsmath,amssymb} 
\usepackage{color}
\usepackage[width=122mm,left=12mm,paperwidth=146mm,height=193mm,top=12mm,paperheight=217mm]{geometry}

\usepackage{booktabs}
\usepackage{pgfplots}
\pgfplotsset{compat=1.14}
\usetikzlibrary{shadows}
\usetikzlibrary{calc}
\usetikzlibrary{spy}
\usetikzlibrary{arrows}
\usetikzlibrary{positioning}
\usepackage{subcaption}
\usepackage[pagebackref=true,breaklinks=true,letterpaper=true,colorlinks,bookmarks=false]{hyperref}

\begin{document}

\newcommand{\introduce}[1]{\emph{#1}\index{#1}} 
\newcommand{\reuse}[1]{\emph{#1}\index{#1}} 

\newcommand{\etal}{\textit{et al}. }

\pagestyle{headings}
\mainmatter
\def \ECCV18SubNumber{1360}  

\title{Surface Defect Classification in Real-Time Using Convolutional Neural Networks} 
\titlerunning{Defect Classification in Real-Time using CNNs}


\author{Selim Arikan\inst{1}, Kiran Varanasi\inst{2} \and
Didier Stricker\inst{2}}

\institute{Technical University of Kaiserslautern,
\email{selimarikan@gmail.com},
\and
German Research Center for Artificial Intelligence (DFKI)
\email{Kiran.Varanasi@dfki.de}, \email{Didier.Stricker@dfki.de}}

\maketitle

\begin{abstract}
Surface inspection systems are an important application domain for computer vision, as they are used for defect detection and classification in the manufacturing industry. Existing systems use hand-crafted features which require extensive domain knowledge to create. Even though Convolutional neural networks (CNNs) have proven successful in many large-scale challenges, industrial inspection systems have yet barely realized their potential due to two significant challenges: real-time processing speed requirements and specialized narrow domain-specific datasets which are sometimes limited in size. In this paper, we propose CNN models that are specifically designed to handle capacity and real-time speed requirements of surface inspection systems. To train and evaluate our network models, we created a surface image dataset containing more than 22000 labeled images with many types of surface materials and achieved 98.0\% accuracy in binary defect classification. To solve the class imbalance problem in our datasets, we introduce neural data augmentation methods which are also applicable to similar domains that suffer from the same problem. Our results show that deep learning based methods are feasible to be used in surface inspection systems and outperform traditional methods in accuracy and inference time by considerable margins.

\end{abstract}

\section{Introduction}
Web manufacturing occupies a large portion of the industry around the globe. A \introduce{web} is a continuously moving flat material such as foil, metal, paper, textile or plastic film. In order to have a production line that can yield high quality and low defect rate products to satisfy the customer demands, web manufacturers use surface inspection systems to check the quality of webs. A \introduce{defect} can be defined as "any undesired outcome in the end-product that reduces the customer satisfaction". Defects can be introduced into the system at any stage and by any system component. 

Surface inspection systems usually utilize a three-tiered machine vision based architecture. First (1) defects are detected and localized, then (2) image is segmented and features of the defect are extracted and (3) the defect is classified. 
\begin{figure}[t]
	\centering
	\begin{subfigure}[t]{0.14\textwidth}
	    \centering
	    \includegraphics[width=\linewidth, height=\linewidth]{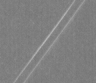}
	    \caption{Film}
	\end{subfigure}
	\begin{subfigure}[t]{0.14\textwidth}
	    \centering
	    \includegraphics[width=\linewidth, height=\linewidth]{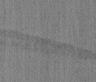}
	    \caption{Foil}
	\end{subfigure}
	\begin{subfigure}[t]{0.14\textwidth}
	    \centering
	    \includegraphics[width=\linewidth, height=\linewidth]{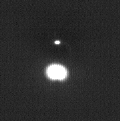}
	    \caption{Glass}
	\end{subfigure}
	\begin{subfigure}[t]{0.14\textwidth}
	    \centering
	    \includegraphics[width=\linewidth, height=\linewidth]{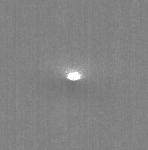}
	    \caption{Coated metal}
	\end{subfigure}
	\begin{subfigure}[t]{0.14\textwidth}
	    \centering
	    \includegraphics[width=\linewidth, height=\linewidth]{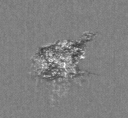}
	    \caption{Steel}
	\end{subfigure}
	\begin{subfigure}[t]{0.14\textwidth}
	    \centering
	    \includegraphics[width=\linewidth, height=\linewidth]{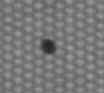}
	    \caption{Non-woven}
	\end{subfigure}
    
    \caption{Various defects from different types of surfaces. In simple cases like (a) and (d), defects can be extracted easily. Advanced cases like (b), (e) and (f) make classification a complex task}
    \label{fig:defectSampleDiagram}
\end{figure}
Traditional systems rely on sophisticated models employing hand-crafted features created with apriori knowledge and high amount of domain expertise. Furthermore, traditional methods are mostly regarded as trademarks, therefore they remain private and difficult to test. 
Even though advancements in production techniques provide higher quality end products, the introduction of defects cannot be prevented completely as they originate from physical processes. Sample defect images from various materials can be seen in Figure \ref{fig:defectSampleDiagram}.

Recently, Convolutional Neural Networks (CNNs) have pushed the envelope of classification performance in visual recognition problems with advancements in network architectures, applications of new ideas and availability of very large datasets which are trainable as a result of improvements in parallel processing \cite{krizhevsky_imagenet_2012} \cite{simonyan_very_2014} \cite{He2015} \cite{szegedy_going_2015}. While human-level performance has been surpassed in generic image classification tasks with CNNs \cite{He2015a}, industrial surface inspection systems barely utilized this potential because web manufacturing systems require methods that can handle classification in real-time.
In this paper, we introduce a defect detection approach for surface inspection systems with a binary classifier using a newly created CNN architecture. 
Our CNN models employ recent developments in deep learning research and designed with surface inspection characteristics and requirements in mind. 


\subsection{Contributions}

\begin{enumerate}
    
    \item We created neural network models from the ground up, designed for surface inspection systems by considering their capacity and very strict requirements in inference time. Our main model achieves the state-of-the-art performance with 99.7\% accuracy on a standard surface dataset and 99.5\% Top-1 accuracy in public multi-class defect dataset (NEU) \cite{Song2013}. 
    
    \item We propose a methodology for collecting a general surface defect dataset and validating the dataset by domain experts for the purpose of training CNN models. 
    
    \item We propose new data augmentation methods to be used with surface images that can improve the testing accuracy by 3.5\%. Our proposed methods employ multiple neural augmentation techniques to generate realistic defect and non-defect material images based on individual class characteristics.
    
    \item Along with the publication of the paper, we will share our deep learning suite along with our models so that other researchers can benefit from our methods, for surface inspection systems as well as for defect classification in other narrow domains.

\end{enumerate}

\section{Background and related work}
\subsection{Convolutional Neural Network Architectures}
The typical structure of convolutional neural networks (CNNs) which arrange convolutional, pooling and fully-connected layers sequentially, started to become standardized after LeNet by LeCun \etal \cite{LeCun}. In 2012, Krizhevsky \etal made the breakthrough with \introduce{AlexNet} \cite{krizhevsky_imagenet_2012} in object recognition by winning the ILSVRC first time with a CNN.

Inspired by a neuroscience model of the primate visual cortex, \introduce{GoogLeNet} from Szegedy \etal used a complex network architecture which used the inputs at multiple levels to extract features using \textit{Inception} modules \cite{szegedy_going_2015}. 

The most common way to increase the performance of a CNNs is increasing their capacity with the cost of increased computational resource usage \cite{szegedy_going_2015}. He \etal has shown that increased depth creates a degradation problem that saturates accuracy of networks without any overfitting \cite{He2015}. To address the problem, He \etal introduced \introduce{deep residual learning} with 152-layer \introduce{ResNet} network with skip-connections between layers \cite{He2015}. ResNet reduced the error rate in ILSVRC down to $3.57\%$ and surpassed human recognition ability \cite{He2015}.

\subsection{Surface Inspection and Defect Detection}

Surface inspection and the defect detection problem can be generalized into the combination of feature extraction and classification problems. 
Even though CNNs are widely used in object detection and image classification tasks, industrial surface inspection systems barely utilize this potential.

Recently Masci \etal introduced Max-Pooling CNN model approach for supervised steel defect classification \cite{Masci2012}. 

On a different approach, Soukup and Huber-M{\"o}rk trained a CNN with stereo imaging to detect steel surface defects \cite{Soukup}. But the stereo acquisition method limits the application and cause the inference speed of this approach to be slow. 

Ke \etal tried using a CNN-based defect recognition in banknote images \cite{Ke2016}. Even though the CNN performs better than traditional methods in results, study of the single type of (circular) defect, limits the usage in similar problems.

Faghih-Roohi \etal used deep learning approaches with multiple CNN models to detect and classify rail surface defects and achieved 92\% accuracy with 5 classes of defects \cite{Faghih-Roohi}. 

Park \etal had a more holistic approach into surface inspection systems with their CNN-based system for surface defect inspection \cite{Park2016}. Park \etal shows that even though CNN-based classifiers perform better than traditional methods with 92\% accuracy, the inference time of $217ms$ is inferior to traditional methods \cite{Park2016}.

Weimer \etal used multiple CNN models to automate the feature extraction in inspection systems \cite{Weimer2016}. Even though results show remarkable accuracy on textured images, they only used the simple circular and linear type of defects but not complex defect types.

Recently Kim \etal applied transfer learning by using the VGG net that is trained on ImageNet \cite{Russakovsky2015} dataset \cite{Kim2017}.

Zhou \etal created a custom CNN model to classify 6 different types of steel defects \cite{Zhou2017}. In contrast, our method is more accurate, converges faster and focuses on not just steel but many materials to achieve surface invariance.

\subsection{Data Augmentation}
Krizhevsky \etal introduced several data augmentation techniques to artificially increase the dataset size using label-preserving transformations  \cite{krizhevsky_imagenet_2012}. 
 
To have more variety in data, rather than only modifying the images, we would like to create new samples to expand our datasets. Recently Goodfellow \etal introduced \introduce{Generative Adversarial Networks} (GANs) to use neural networks to generate new samples using adversarial training \cite{goodfellow_explaining_2014} \cite{goodfellow_generative_2014-1}. 

Using the conditional GANs, compared to other domain-specific methods, Isola \etal introduced a general-purpose paired image translation method also known as \introduce{pix2pix} \cite{Isola}. 

Because obtaining paired image data is expensive and difficult, Zhu \etal introduced a cycle-consistent adversarial network architecture called \introduce{CycleGAN} for unpaired image translation problems \cite{Zhu}. 

With the advancements of synthetic image generation, it has become a common practice to use generated images in training neural networks to avoid the high cost of creating large datasets with real images. 
Shrivastava \etal introduced an improved approach to image generation with Simulated+Unsupervised learning (SimGAN) which uses synthetic images rather than random vectors as inputs to their GAN \cite{Shrivastava}. By using a self-regularization term and a local adversarial loss, SimGAN converts synthetic renderings into realistic images without using any labeled data \cite{Shrivastava}. Their method is able to achieve local changes without altering the global structure of the image. In contrast, we propose a data augmentation method for altering global scene composition in the image. 

\section{Methodology}
\subsection{Dataset Creation}\label{sect:datasetCollection}
In order to create high-quality datasets, we established a three-step methodology, which we validated in a real world industrial setting for surface inspection. 
\subsubsection{Aggregation}
Firstly, we collect raw images, other small datasets and individual samples from many sources. All of the collected images are converted into same format and channel/bitrate for standardization. Finally the images are divided into 'defect' and 'non-defect' classes.

\subsubsection{Cleansing} From the aggregated raw images, we delete the invalid images such as non-web surfaces, empty material edges and faulty samples that are completely black or white. If cleansing is not applied, models would try to use invalid samples and consequently, learn wrong features. 

\subsubsection{Balancing} Finally we balance our datasets by several criteria in order to correctly represent different materials and defect types. It is important to contain many types of web materials in the dataset but having a balance between different samples of different materials is equally important. Same principle also applies for defect types. Defect-free samples and some types of defects dominate the occurrence in the inspection systems. However the diversity and severity of various defects needs to be taken into account. A defect may occur only 1\% of the time but can have severe consequences if not detected. An example for this in healthcare is the detection of cancer patients. 
If balancing is not done, some types of materials and defects would be under or overrepresented in the dataset, causing an imbalance which would affect the learning and prevent achieving high success rates.


\subsection{Data Augmentation}


We divide our data augmentation techniques into two phases. In the first phase, we employed offline affine and generative label-preserving methods where dataset size is directly increased by creating new samples from existing images. To realize this, we used a selection of GANs (pix2pix, CycleGAN) with paired and unpaired image samples to generate new defect and non-defect images respectively \cite{Isola} \cite{Zhu}. 

In the second phase we used online affine methods where transformations are label-preserving and applied via an integrated operation of the framework at the mini-batch selection. 
We also have a process of validating the applicability of the synthetic images by asking domain experts to choose between real and synthetic images in a survey. 
More details about our data augmentation process can be found on Section \ref{section:Aug}.

\subsection{Network Architecture} \label{section:SurfNet}

\tikzstyle{convn}=[draw, fill=green!30, minimum size=1em]
\tikzstyle{conv1}=[draw, fill=green!10, minimum size=1em]
\tikzstyle{pool}=[draw, fill=blue!10, minimum size=1em]
\tikzstyle{fc}=[draw, fill=orange!20, minimum size=1em]

\begin{figure}[h]
\begin{tabular}[c]{llr}
    \begin{subfigure}[t]{0.22\textwidth}
    \hfill
    \begin{tikzpicture}[node distance=0cm, auto, >=latex', scale=0.6, every node/.style={scale=0.6}]
        \node [convn] (in) {\footnotesize Conv $5\times5$ / 2};
        \node [conv1] (iA) [below=0.25cm of in] {\footnotesize Conv $1\times1$ / 1};
        \node [convn] (i2) [below=0.25cm of iA] {\footnotesize Conv $5\times5$ / 2};
        \node [conv1] (iB) [below=0.25cm of i2] {\footnotesize Conv $1\times1$ / 1};
        \node [convn] (i3) [below=0.25cm of iB] {\footnotesize Conv $5\times5$ / 2};
        \node [conv1] (iC) [below=0.25cm of i3] {\footnotesize Conv $1\times1$ / 1};
        \node [conv1] (i4) [below=0.35cm of iC] {\footnotesize Conv $1\times1$ / 1};
        \node [conv1] (i5) [below=0.25cm of i4] {\footnotesize Conv $1\times1$ / 1};
        \node [conv1] (i6) [below=0.25cm of i5] {\footnotesize Conv $1\times1$ / 1};
        
        \node [fc] (linear) [below=0.5cm of i6] {\footnotesize FC n:2};
        
        \node [draw, fill=red!10, minimum size=1em] (input) [above=0.5cm of in] {\footnotesize Input image};
        
        \path[->] (input) edge node {} (in);
        \path[->] (in) edge node {} (iA);
        
        \path(in) -- coordinate (branch1) (iA);
        \path(iA) -- coordinate (branch2) (i2);
        \draw[->] (branch1) to [out=0, in=0, min distance=2.5cm] (branch2);

        \path[->] (iA) edge node {} (i2);
        \path[->] (i2) edge node {} (iB);
        
        \path(i2) -- coordinate (branch3) (iB);
        \path(iB) -- coordinate (branch4) (i3);
        \draw[->] (branch3) to [out=0, in=0, min distance=2.5cm] (branch4);
        
        \path[->] (iB) edge node {} (i3);
        \path[->] (i3) edge node {} (iC);
        
        \path(i3) -- coordinate (branch5) (iC);
        \path(iC) -- coordinate (branch6) (i4);
        \draw[->] (branch5) to [out=0, in=0, min distance=2.5cm] (branch6);
        
        \path[->] (iC) edge node {} (i4);
        
        \path(i4) -- coordinate (branch7) (i5);
        \draw[->] (branch6) to [out=0, in=0, min distance=2.5cm] (branch7);
        
        \path[->] (i4) edge node {} (i5);
        
        \path(i5) -- coordinate (branch8) (i6);
        \draw[->] (branch7) to [out=0, in=0, min distance=2.5cm] (branch8);
        
        \path[->] (i5) edge node {} (i6);
        
        \path(i6) -- coordinate (branch9) (linear);
        \draw[->] (branch8) to [out=0, in=0, min distance=2.5cm] (branch9);
        
        \path[->] (i6) edge node {} (linear);
        
    \end{tikzpicture}
    \caption{SurfNet}
    \label{fig:dsmnlv4}
    \end{subfigure} &
    
    \begin{subfigure}[t]{0.22\textwidth}
    \centering
    \begin{tikzpicture}[node distance=0.2cm, auto, >=latex', scale=0.6, every node/.style={scale=0.6}]
        \node [convn] (in) {\footnotesize Conv $3\times3$ / 2};
        \node [conv1] (i2) [below=0.25cm of in] {\footnotesize Conv $1\times1$ / 1};
        \node [convn] (i3) [below=0.25cm of i2] {\footnotesize Conv $3\times3$ / 2};
        
        \node [fc] (linear) [below=0.5cm of i3] {\footnotesize FC n:2};
        
        \node [draw, fill=red!10, minimum size=1em] (input) [above=0.5cm of in] {\footnotesize Input image};
        
        \path[->] (input) edge node {} (in);
        \path[->] (in) edge node {} (i2);
        \path[->] (i2) edge node {} (i3);
        \path[->] (i3) edge node {} (linear);
        
    \end{tikzpicture}
    \caption{FastInf}
    \label{fig:fastinf}
    \end{subfigure} &
   \begin{subfigure}[t]{0.45\textwidth}
    \centering
    \begin{tikzpicture}[node distance=1.5cm, auto, >=latex', scale=0.6, every node/.style={scale=0.6}]
        
        
        \node [convn] (hpin) {\footnotesize Conv $3\times3$ / 2};
        \node [conv1] (hpiA) [below=0.25cm of hpin] {\footnotesize Conv $1\times1$ / 1};
        \node [convn] (hpi2) [below=0.25cm of hpiA] {\footnotesize Conv $3\times3$ / 2};
        \node [conv1] (hpiB) [below=0.25cm of hpi2] {\footnotesize Conv $1\times1$ / 1};
        \node [convn] (hpi3) [below=0.25cm of hpiB] {\footnotesize Conv $3\times3$ / 2};
        
        \node [convn] (bpin) [right=0.7cm of hpin] {\footnotesize Conv $5\times5$ / 2};
        \node [conv1] (bpiA) [below=0.25cm of bpin] {\footnotesize Conv $1\times1$ / 1};
        \node [convn] (bpi2) [below=0.25cm of bpiA] {\footnotesize Conv $5\times5$ / 2};
        \node [conv1] (bpiB) [below=0.25cm of bpi2] {\footnotesize Conv $1\times1$ / 1};
        \node [convn] (bpi3) [below=0.25cm of bpiB] {\footnotesize Conv $3\times3$ / 2};
        
        \node [convn] (lpin) [right=0.7cm of bpin] {\footnotesize Conv $3\times3$ / 2};
        \node [conv1] (lpiA) [below=0.4cm of lpin] {\footnotesize Conv $1\times1$ / 1};
        \node [pool] (lpi2) [below=0.4cm of lpiA] {\footnotesize MaxPool $2\times2$ / 2};
        \node [convn] (lpi3) [below=0.4cm of lpi2] {\footnotesize Conv $3\times3$ / 2};
        
        \node [draw, fill=red!10, minimum size=1em] (input) [above=0.5cm of bpin] {\footnotesize Input image};
        
        \node [conv1] (api3) [below=0.5cm of bpi3] {\footnotesize Conv $1\times1$ / 1};
        \node [fc] (linear) [below=0.5cm of api3] {\footnotesize FC n:2};
        
        \path[->] (input) edge node {} (hpin);
        \path[->] (input) edge node {} (bpin);
        \path[->] (input) edge node {} (lpin);
        
        \path[->] (hpin) edge node {} (hpiA);
        \path[->] (hpiA) edge node {} (hpi2);
        \path[->] (hpi2) edge node {} (hpiB);
        \path[->] (hpiB) edge node {} (hpi3);
        \path[->] (hpi3) edge node {} (api3);
        
        \path[->] (bpin) edge node {} (bpiA);
        \path[->] (bpiA) edge node {} (bpi2);
        \path[->] (bpi2) edge node {} (bpiB);
        \path[->] (bpiB) edge node {} (bpi3);
        \path[->] (bpi3) edge node {} (api3);
        
        \path[->] (lpin) edge node {} (lpiA);
        \path[->] (lpiA) edge node {} (lpi2);
        \path[->] (lpi2) edge node {} (lpi3);
        \path[->] (lpi3) edge node {} (api3);
        
        \path[->] (api3) edge node {} (linear);
        
    \end{tikzpicture}
    \caption{MultiVis}
    \label{fig:channelNet}
    \end{subfigure} 
    \end{tabular}
    \caption{Our network models: curved arrows indicate residual skip-connections. Stride value is shown with last number in each block.}
\end{figure}
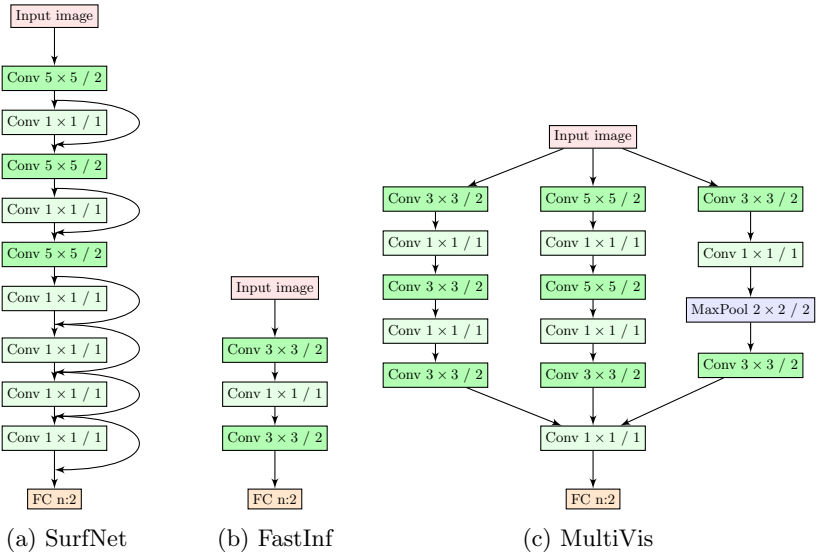

Since web manufacturing lines operate at very fast speeds, surface inspection systems have to work quickly (\textasciitilde 100 images/second, 10m/s) as well. Many of the famous network models aim for the highest classification accuracy but completely ignore the costs in inference time. Because of their long inference time and consequently higher computational resource and power consumption, practical and real-time applications of the common network models are extremely limited \cite{Canziani2016}. The main goal of our architecture is to capture defect information globally with the highest accuracy and as fast as possible to satisfy real-time requirements of surface inspection systems. 

As seen in Figure \ref{fig:dsmnlv4}, our main model (SurfNet) is generally inspired by the ideas of VGG net configurations \cite{simonyan_very_2014} and the idea of residual learning \cite{He2015}. 

We used ten layers with first nine are convolutional for feature extraction and the last is a fully-connected layer for the classification. All of the convolutional blocks contain a batch normalization \cite{Ioffe2015} and an activation function layer. 

We adopted \introduce{parametric rectified linear unit} (PReLU) from He \etal over ReLU for its benefits and ability to be optimized while training \cite{He2015a}. Following the practice in \cite{Ioffe2015}, we did not use any \introduce{dropout} \cite{Hinton} layers together with batch normalization \cite{Ioffe2015}. To gain more speed, spatial size reduction is handled by convolution operations ($stride = 2$), therefore no pooling layers are used. 

We employed two types of convolutional layers. The first type is used to extract features and it uses larger $5\times5$ receptive fields (i.e. kernel size), has $6$ pixels of padding and performs downsampling directly with a stride value of $2$. Padding values of the convolutional layers are selected appropriately so that spatial resolution is not altered by the kernel size. 
Second type of convolutional layer we use utilizes $1\times1$ receptive fields without padding and  downsampling ($stride=1$). At first glance, using $1\times1$ kernel may seem redundant but it serves many purposes: following the practice in Szegedy \etal, they are used as a dimension reduction method and to provide additional nonlinearity with their activation functions \cite{szegedy_going_2015}. Lastly, we used $1\times1$ convolutional layers to apply residual learning \cite{He2015} by adding shortcut connections over them.

We followed the practice in \cite{He2015} \cite{szegedy_going_2015} and did not use fully-connected layers because the benefit was negligible whereas the computational cost was much higher. 

\subsection{Training Process}

We used mini-batch training method with RMSProp optimizer and a negative log-likelihood loss function. We set the batch size to 10 and learning rate to $1.10^{-4}$. For regularization, weight decay ($L_2$) method is used with $1.10^{-1}$ multiplier. Learning rate (LR) is adjusted with a step function decreasing the LR every 3 epochs with a multiplier value of $0.8$. We follow the practice in \cite{He2015} to initialize the weights of our convolutional and batch normalization layers. The training and all of the experiments are done on a standard Windows desktop computer using an NVIDIA GTX 1080 Ti GPU with \introduce{PyTorch} \cite{PyTorch} deep-learning framework up to 100 epochs depending on the configuration and dataset. 

For each mini-batch, pre-processing operations applied to resize and crop the images to $128\times128$ pixels. For resizing and cropping, we followed the practice in \cite{simonyan_very_2014}.

\subsection{Testing Process}

Our test sets contain images cropped to their defective regions provided by domain experts in order to test the classification performance without performing localization. First, the image is scaled to 128 pixels with respect to its shorter side, then the center $128\times128$ patch is cropped for testing. 
Contrary to the training process, random horizontal flipping is not applied in testing.

\subsection{Alternative Network Configurations}

We experimented with many ideas and advancements in deep learning research. For a qualitative analysis of the different network models for the problem domain of surface inspection, we created two different network models and compared them with the original SurfNet model. Our first alternative model is aimed at achieving minimum inference time and our second model is aimed at parallel processing at multiple scales.

\subsubsection{Minimum Inference Time - FastInf} \label{section:FastInf}

The biggest trade-off in designing a CNN model is to achieve a balance between accuracy and inference time. Since surface inspection systems require real-time detection to keep up with the manufacturing speed requirements, we tried to achieve the best inference time while keeping the accuracy better than traditional classification methods. Rather than increasing the depth, we made the model wide with 1024 convolution channels. As a result, the \introduce{FastInf} configuration can achieve $0.23ms$ inference time per image in binary classification with a desktop computer using an NVIDIA GTX 1080 Ti.  Model can be seen in Figure \ref{fig:fastinf}.

\subsubsection{Parallel Multi-Scale Processing - MultiVis} \label{section:MultiVis}

The primate visual cortex recognizes objects on multiple levels and scales. To imitate this process, we designed a model which has 3 parallel lanes with each of them processing the input in different scales. The similar idea is used by Szegedy \etal \cite{szegedy_going_2015} but as individual blocks known as Inception layers. Rather than using blocks, we used 3 parallel optimized fast inference network models to improve the capacity of the model.

After the input image is processed in 3 different parallel channels, results are concatenated and provided as input to the last $1\times1$ convolutional layer. Model can be seen in Figure \ref{fig:channelNet}.

\section{Data augmentation} \label{section:Aug}

The perfect method to train a CNN is to supply it with a lot of labeled data. Even though many general-purpose datasets are easily available, acquiring domain-specific labeled data is difficult and often the data is not evenly distributed between classes. The class imbalance is expected because occurrences for each class do not have the same frequency. Having insufficient amount of data potentially leads networks to overfit and prevents generalization. When obtaining more real data is not possible or expensive, data augmentation is the ideal way to increase the size of the datasets by creating new samples using various methods.

To evaluate the results of our augmentation methods, we created quantitative, qualitative and user A/B testing experiments. We have seen that, with our proposed method, test accuracies of our models increased by up to 3.5\%. Also, user experiments show us that, even by domain experts, created images are almost indistinguishable from real samples.

\subsection{Classical Methods}

Affine transformations such as rotation, scaling, mirroring and shearing are the simplest methods that are performed as the first choice almost in any set because of their simple and quick implementation \cite{Howard2013}. 

\subsection{Generative Neural Methods}

Even though the simple techniques derive new samples from existing images, they do not create unique images. Generating images using models is a well-studied topic that has beneficial uses in many fields. 

\subsubsection{Paired Image Translation with Conditional Adversarial Networks}

Isola \etal introduced a paired image translation method using conditional adversarial networks which is known as \introduce{pix2pix} to create a general-purpose solution to image-to-image translation tasks \cite{Isola}.

In surface inspection, because defects originate from physical sources especially from imperfections in production processes, they are localized and usually appear clustered in regions. Therefore we used the label-to-image conversion capability of pix2pix for generating synthetic defect images. 
Using label-to-image translation allowed us to generate as many defect images as necessary. We automated the label generation process and we can define any defect region with any shape we want. Furthermore, custom label creation can be used by domain experts to tailor image generation process to reflect physical characteristics of the given web material, manufacturer, equipment and imperfection profiles to model the system better. 
Lastly, by creating necessary samples for the underrepresented defect classes, our automatic label generation method can be used to solve the class imbalance problem.

\subsubsection{Unpaired Image Translation with Cycle-consistent Adversarial Networks}

Not only acquiring paired data is difficult and often expensive, but also the paired datasets are rare and considerably smaller in size compared to standard datasets. Zhu \etal offers a general-purpose solution to this problem with an unpaired image-to-image translation method using cycle-consistent adversarial networks that is called \introduce{CycleGAN} \cite{Zhu}. 

\begin{figure}[h]
    \centering
    \begin{subfigure}[t, sub]{0.25\textwidth}
        \centering
        \includegraphics[width=\linewidth]{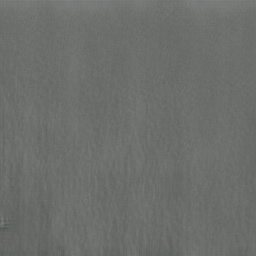}
        \caption{Synthetic 1}
    \end{subfigure}
    \begin{subfigure}[t, sub]{0.25\textwidth}
        \centering
        \includegraphics[width=\linewidth]{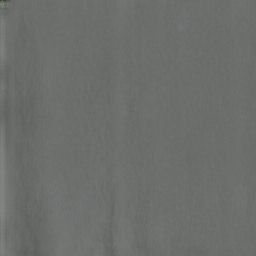}
        \caption{Synthetic 2}
    \end{subfigure}
    \begin{subfigure}[t, sub]{0.25\textwidth}
        \centering
        \includegraphics[width=\linewidth]{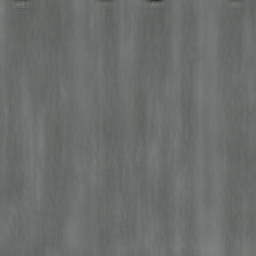}
        \caption{Synthetic 3}
    \end{subfigure}
    \caption{We apply the unpaired image-to-image translation to create non-defect images using SetA and SetB datasets}
    \label{fig:augcgResults}
\end{figure}

Unlike defect images with local characteristics, non-defect material images do not have specific features that we can use directly. Non-defect images have global variations which are caused by natural processes, inspection system characteristics and web material features. 
Therefore, non-defect images require an augmentation method which can capture their global characteristics. We, therefore, used style transfer capability of CycleGAN in augmenting non-defect images of our datasets with the goal of mixing and matching variations in different sets to create unique samples. Sample synthetic images can be seen in Figure \ref{fig:augcgResults}.

\begin{figure}
	\centering
	\begin{subfigure}[t, sub]{0.25\textwidth}
	    \centering
	    \includegraphics[width=\linewidth]{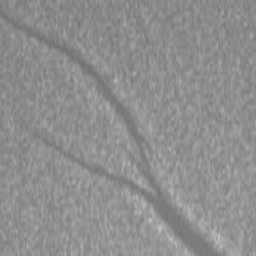}
	    \caption{Real defect in $B$}
	\end{subfigure}
    \begin{subfigure}[t, sub]{0.25\textwidth}
        \centering
	    \includegraphics[width=\linewidth]{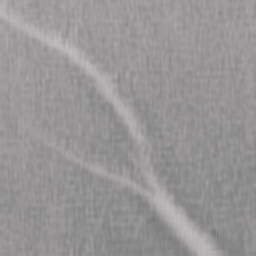}
	    \caption{Synthetic}
	\end{subfigure}
	\begin{subfigure}[t, sub]{0.25\textwidth}
        \centering
	    \includegraphics[width=\linewidth]{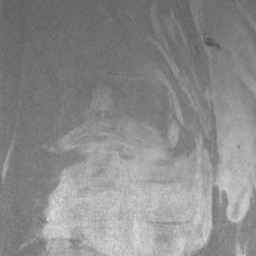}
	    \caption{Real defect in $A$}
	\end{subfigure}
	\caption{Using unpaired image generation for defects. It can be observed that network has learned to translate the bright and smooth defect characteristics of (c) to dark and high-contrast defect characteristics of (a), resulting image (b) with bright and smoother defect pixels}
    \label{fig:augCGDefectPairs}
\end{figure}

In addition to non-defect image generation, we tried using unpaired image generation technique on defect images with the ambition of creating unique defects by learning features and style of defects on different surfaces. Using unpaired technique helps us bridge the gap between different characteristics of various defect classes. The result can be seen in Figure \ref{fig:augCGDefectPairs}.

\subsection{Class Imbalance}
\introduce{Class imbalance} is a common problem in machine learning where some classes in a dataset are underrepresented (i.e. have fewer samples than other classes). Reasons for class imbalance can vary but commonly the problem arises when it is not possible or very difficult to gather more samples for a specific class such as fraud cases in banking and data of cancer patients in healthcare. 

In surface inspection, defects are encountered in approximately 0.1\% of the samples. On the contrary, since the inspection systems are aimed towards detecting defects and capturing images at extremely fast speeds, only the relevant images containing defects can be saved and the rest are discarded. This results in having almost no defect-free images in captured data. 

To overcome this problem, we used classical as well as neural data augmentation techniques to derive and generate more samples. More details on image creation can be found in Section \ref{subsec:selectedAug}.

\subsection{Selected Methods} \label{subsec:selectedAug}

\begin{figure*}[ht!]
\begin{center}
    \begin{subfigure}[t, sub]{0.19\textwidth}
        \centering
        \includegraphics[width=\linewidth]{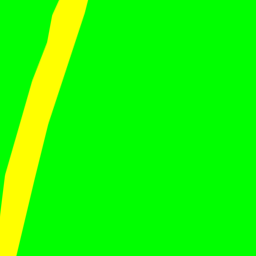}
        \caption{Label 1}
    \end{subfigure}
    \begin{subfigure}[t, sub]{0.19\textwidth}
        \centering
        \includegraphics[width=\linewidth]{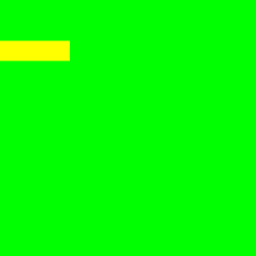}
        \caption{Label 3}
    \end{subfigure}
    \begin{subfigure}[t, sub]{0.19\textwidth}
        \centering
        \includegraphics[width=\linewidth]{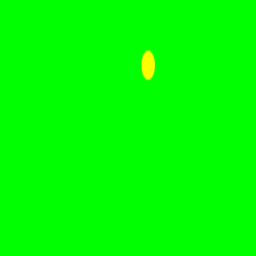}
        \caption{Label 4}
    \end{subfigure}
    \begin{subfigure}[t, sub]{0.19\textwidth}
        \centering
        \includegraphics[width=\linewidth]{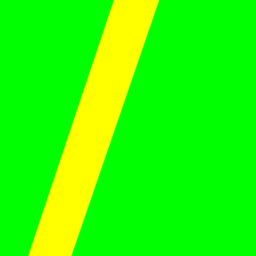}
        \caption{Label 5}
    \end{subfigure}
    \begin{subfigure}[t, sub]{0.19\textwidth}
        \centering
        \includegraphics[width=\linewidth]{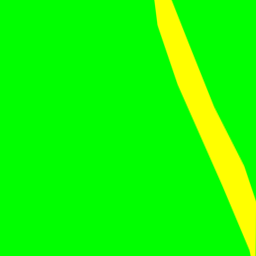}
        \caption{Label 6}
    \end{subfigure}
 
    \begin{subfigure}[t, sub]{0.19\textwidth}
        \centering
        \includegraphics[width=\linewidth]{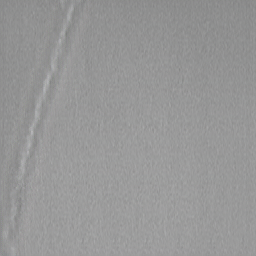}
        \caption{Synthetic 1}
    \end{subfigure}
    \begin{subfigure}[t, sub]{0.19\textwidth}
        \centering
        \includegraphics[width=\linewidth]{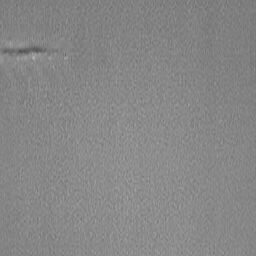}
        \caption{Synthetic 3}
    \end{subfigure}
        \begin{subfigure}[t, sub]{0.19\textwidth}
        \centering
        \includegraphics[width=\linewidth]{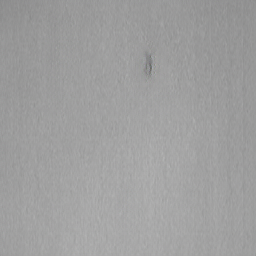}
        \caption{Synthetic 4}
    \end{subfigure}
    \begin{subfigure}[t, sub]{0.19\textwidth}
        \centering
        \includegraphics[width=\linewidth]{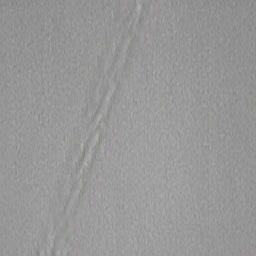}
        \caption{Synthetic 5}
    \end{subfigure}
    \begin{subfigure}[t, sub]{0.19\textwidth}
        \centering
        \includegraphics[width=\linewidth]{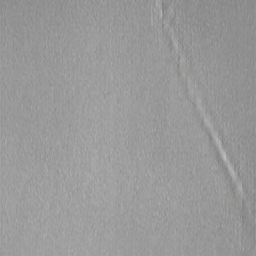}
        \caption{Synthetic 6}
    \end{subfigure}
    
\end{center}
   \caption{We apply label-to-image translation for defect image generation. We translate the label images with green (defect-free) background and yellow regions (defective areas), to synthetic defect images which are visually realistic. }
    \label{fig:augp2pResults}
\end{figure*}

For classical methods, we did not use any color based data augmentation methods since our images are monochromatic. We used rotations at offline stage and used horizontal mirroring and scale-cropping randomly during runtime at image preprocessing step where mini-batches are prepared. 

For neural augmentation methods, we used pix2pix and CycleGAN for defect and non-defect classes respectively considering class characteristics and capabilities of generative methods. 

We trained pix2pix model with default parameters and "UNet 256" generator with batch normalization for 200 epochs. We used label-to-image translation to create synthetic defect images. In order to create label images, we marked defective regions positioned in specific locations and shapes in the image (e.g. closer to edges, non-rectangular concave shapes and diagonal stripes). Some examples of label images and their generated images can be seen in Figure \ref{fig:augp2pResults}. Furthermore, to automatize the process, we created a label generator which creates label images with parametric shapes in various sizes and in variable quantities. Automatically generated label image samples can be found in supplementary material.

For non-defect image generation, we trained CycleGAN model using default parameters without dropout for 200 epochs. We used the non-defect material image classes of the two sets (SetA and SetB) as $A$ and $B$ sets for the network to learn the translation $G: A \rightarrow B$ between them. Sample results of the unpaired image translation between different datasets can be seen in Figure \ref{fig:augcgResults} and \ref{fig:augCGDefectPairs}. More samples can be found in supplementary material.

\subsection{Evaluation}

Evaluating the quality of synthetic images is a difficult problem because there is not a simple way to determine whether an image looks real or not. Neural augmentation methods such as GANs also do not contain an objective function to determine the "authenticity" of the images. For quantitative evaluation, we created a basic A/B testing to be presented to domain experts for them to pick which image they think in the given A/B pair is synthetic. From the provided 10 image pairs, domain experts were able to choose the synthetic images with only 60\% accuracy which indicates synthetic images are realistic enough for test takers to convince them. More details about our quality evaluation test can be found in supplementary material.

\section{Results}
\subsection{Performance Evaluation Metrics}
For evaluating the performance of the models, we used the standard classification metrics, specifically, \introduce{accuracy}, \introduce{precision}, \introduce{recall}(sensitivity) and \introduce{specificity} \cite{Sokolova2009}. 




    


\subsection{Datasets}
For training and experiments, we used five datasets in which two of them (SetA and SetB) are being used as a standard to evaluate the accuracy of currently used C5.0 classifiers. The third dataset SetAB combines SetA and SetB. 
We created the fourth dataset (SurfMix) using our dataset collection method explained in Section \ref{sect:datasetCollection} with the goal of creating a surface-type-invariant binary classifier to be used in any system with any type of surface. Numerical details about the internal datasets can be seen in Table \ref{tab:datasetNumbers}.

Industrial datasets are unfortunately almost always private and hinder reproducibility of the methods. Because of that, as the fifth, we also tested our methods using the public NEU multi-class steel defect dataset \cite{Song2013}. NEU consists 6 classes (crazing, inclusion, patches, pitted surface, rolled-in-scale and scratches) each containing 300 images. As the NEU dataset does not provide separate train and test samples, we used 10-fold cross-validation for training and test.

\begin{table}[h]
\centering
\begin{tabular}{@{}lcccc@{}}
\toprule
        & \multicolumn{2}{c}{Non-augmented} & \multicolumn{2}{c}{Augmented} \\ \cmidrule(l){2-3} \cmidrule(l){4-5}
Dataset   & Defect        & Non-defect       & Defect   & Non-defect        \\ \midrule
SetA    & 351       & 256       & 1403  & 1022     \\
SetB    & 147       & 214       & 588   & 856      \\ 
SetAB   & 498       & 470       & 2470  & 2288     \\ 
SurfMix & 8359      & 7173      & 33436 & 28692    \\ 
\bottomrule
\end{tabular}
\caption{Training images in internal datasets. SetA and SetB are standard classifier evaluation sets. SurfMix is our newly created dataset.}
\label{tab:datasetNumbers}
\end{table}

\subsection{Model Comparison}

We achieved state-of-the-art binary classification performance in the datasets with all of our models compared to currently used classifiers. Best accuracy is achieved by SurfNet model resulting 100.0\% in SetA, 97.7\% in SetB and 98.0\% in SurfMix dataset with 1.9 ms inference time. The fastest inference time is achieved by FastInf model with 0.3 ms on 99.3\% accuracy in SetA and 88.6\% accuracy in SetB datasets. Details about the network models can be found in Section \ref{section:SurfNet}.

\begin{table*}[t]
\centering
\begin{tabular}{@{}lccccccc@{}}
\toprule
& \multicolumn{3}{c}{SetA accuracy} & \multicolumn{3}{c}{SetB accuracy} & \\ \cmidrule(l){2-4}  \cmidrule(l){5-7}
Model              & Overall  & Defect & Non-defect & Overall  & Defect & Non-defect & Inference time    \\ \midrule
Traditional (C5.0) & 0.874            & 0.922 & 0.827    & 0.777              & 0.863  & 0.691 & 15.0ms  \\
SurfNet (CNN)      & \textbf{0.993}            & \textbf{0.988} & 1.000    & 0.945              & \textbf{0.977}  & 0.870 & 1.9ms  \\
MultiVis (CNN)     & 0.990            & 0.983 & 1.000    & \textbf{0.957}              & 0.960  & \textbf{0.950} & 3.5ms  \\
FastInf (CNN)      & 0.990            & 0.983 & 1.000    & 0.863              & \textbf{0.988}  & 0.690 & \textbf{0.3ms} \\
\bottomrule
\end{tabular}%
\caption{All of our models achieve better results compared to C5.0 classifier. For a correct comparison, models are trained on non-augmented datasets. Inference time is calculated per image (batch size = 1) and from an average of 10 runs.}
\label{tab:traditionalvsCNN}
\end{table*}

\begin{table}[h]
\centering
\begin{tabular}{@{}lccccccc@{}}
\toprule
Model \phantom{abcdefghjkl} & Top-1 accuracy \phantom{a} & Epochs \phantom{a} & Learning rate \phantom{a} & $L_2$  & Inference time    \\ \midrule
Zhou \etal         & 0.990          & 100    & 0.001         & 0.0005 & \\ 
Zhou \etal         & 0.992          & 300    & 0.001         & 0.0005 & \\ 
Zhou \etal         & 0.993          & 500    & 0.001         & 0.0005 & \\ 
SurfNet (ours)     & \textbf{0.995} & 100    & 0.0007        & 0.02   & 1.9ms  \\
MultiVis (ours)    & 0.984          & 50     & 0.001         & 0.01   & 3.4ms  \\
FastInf (ours)     & 0.944          & 100    & 0.001         & 0.003  & 0.2ms  \\
\bottomrule
\end{tabular}%
\caption{Public NEU dataset results.  Our main model achieve almost perfect result in test accuracy while keeping the real-time performance. Results are average of 10-folds. For experiments, batch size is 10 and image size is $64\times64$. Zhou \etal used $40\times40$ images and batch size of 50. Inference time is per image}
\label{tab:publicResults}
\end{table}




The currently used traditional classifiers we compare to are C5.0 decision trees with over 400 hand-crafted features that require extensive domain knowledge. Even though the effort of developing the traditional classifiers is high, they are not robust to small changes and achieved accuracy rates do not seem to reflect this effort. Comparison of classifiers can be seen in Table \ref{tab:traditionalvsCNN}. 

Our model achieved state-of-the-art Top-1 accuracy in the NEU dataset while keeping real-time inference requirements. Comparisons can be seen in Table \ref{tab:publicResults}. Additional results, hyperparameter evaluations and training graphs can be found in supplementary material.


\subsection{Benefits of Data Augmentation}
We trained our models both on non-augmented and augmented datasets to compare the accuracy values. Methods we use improve the test accuracy up to 3.2\%. Detailed results can be seen in Table \ref{tab:AugEffects}.

\begin{table}[h!]
\centering
\begin{tabular}{@{}lcccccc@{}}
\toprule
Model \phantom{abcd}  & \phantom{a}SetA\phantom{a}  & SetA (Aug)\phantom{a} & \phantom{a}SetB\phantom{a}& SetB (Aug)\phantom{a} & \phantom{a}SurfMix\phantom{a}& SurfMix (Aug) \\ \midrule
SurfNet & 0.993 & 1.000 & 0.945  & 0.977 & 0.980   & 0.983    \\
FastInf & 0.990 & 0.993 & 0.863  & 0.886 & 0.949   & 0.941    \\ \bottomrule
\end{tabular}%
\caption{Test accuracy values after 30 epochs. Our augmentation methods provide up to 3.2\% increase in accuracy. We observe that data augmentation provide more benefits to the network model that has more learning capacity (i.e. SurfNet) compared to the FastInf model. The negligible accuracy difference in augmented SurfMix shows that dataset is already diverse. }
\label{tab:AugEffects} 
\end{table}
\section{Training and Inference Time (Ours vs common models)}
We compare our network model with famous models (ResNet, DenseNet etc.) with respect to training time and inference time. Results show that our model is almost 20 times faster in inference time than the DenseNet model with only 1\% drop in test accuracy. The results can be seen in Table \ref{tab:commonModels}.

\begin{table}[h!]
\centering
\setlength{\textfloatsep}{0.1cm}
\begin{tabular}{@{}lccc@{}}
\toprule
Model  \phantom{abcdef}  & Accuracy & Train time & Inference time \\ \midrule
SurfNet (ours)  & 0.980    &  158m22s      & 1.92ms         \\
ResNet18 & 0.982    &  162m45s      & 11.53ms        \\
DenseNet & 0.990    &  665m50s      & 39.12ms         \\ \bottomrule
\end{tabular}%
\caption{Comparison of our model with common models using SurfMix dataset. ResNet \cite{He2015}, DenseNet\cite{Huang2016}. Trained for 30 epochs using $224\times224$ image size. }
\label{tab:commonModels}
\end{table}
\section{Conclusions}
In this paper, we proposed new convolutional neural network models for binary classification of surface defects and compared it to traditional classifiers. Our model is fast and accurate, making it suitable for deployment in the industrial setting. Contrary to individual design and tailoring of classifiers per surface type, customer, manufacturing line and even per system, we used a surface-invariant and generalized approach. First, we proposed a methodology for data set acquisition in this domain, and created a new surface data set containing images from 23 different actively-used systems with many types of surface materials such as steel, paper, foil, glass, plastic and film. We designed 3 new CNN models from ground-up with practical applicability and real-time requirements of surface inspection systems in focus. We used neural data augmentation methods that are novel in surface inspection domain to solve class imbalance problems in our datasets. Consequently, we tested our network models with five data sets and achieved 98.0\% accuracy in general SurfMix dataset and outperforming standard classifiers in all tests. To further verify our methodology, we tested our models with public NEU dataset and achieved 99.5\% Top-1 accuracy with real-time inference. Thus, our results conclude that CNNs are viable alternatives to standard hand-crafted classifiers in binary classification of surface images. 

Since our solution tackles a general object recognition task, our models and methods are not only applicable for commercial surface inspection systems but also beneficial for other problems with similar domain characteristics (goal of pattern recognition, scarcity of samples and problem of class imbalance). In future work, these models can be extended to handle more complex changes such as in viewpoint, illumination and other domain-specific geometric variability. We share our deep learning suite and network models to encourage such future works in various application domains requiring real-time response.





\clearpage

\bibliographystyle{splncs}
\bibliography{arxiv_arikan}

\begin{thebibliography}{10}

\bibitem{krizhevsky_imagenet_2012}
Krizhevsky, A., Sutskever, I., Hinton, G.E.:
\newblock {ImageNet Classification with Deep Convolutional Neural Networks}.
\newblock In: Advances In Neural Information Processing Systems. (2012)  1--9

\bibitem{simonyan_very_2014}
Simonyan, K., Zisserman, A.:
\newblock {Very Deep Convolutional Networks for Large-Scale Image Recognition}.
\newblock International Conference on Learning Representations (ICRL) (sep
  2015)  1--14

\bibitem{He2015}
He, K., Zhang, X., Ren, S., Sun, J.:
\newblock {Deep Residual Learning for Image Recognition}.
\newblock Technical report (2015)

\bibitem{szegedy_going_2015}
Szegedy, C., Liu, W., Jia, Y., Sermanet, P., Reed, S., Anguelov, D., Erhan, D.,
  Vanhoucke, V., Rabinovich, A., Hill, C., Arbor, A.:
\newblock {Going Deeper with Convolutions}.
\newblock In: Proceedings of the {\{}IEEE{\}} {\{}Conference{\}} on
  {\{}Computer{\}} {\{}Vision{\}} and {\{}Pattern{\}} {\{}Recognition{\}}.
  (2014)  1--9

\bibitem{He2015a}
He, K., Zhang, X., Ren, S., Sun, J.:
\newblock {Delving Deep into Rectifiers: Surpassing Human-Level Performance on
  ImageNet Classification}.
\newblock (feb 2015)

\bibitem{Song2013}
Song, K., Yan, Y.:
\newblock {A noise robust method based on completed local binary patterns for
  hot-rolled steel strip surface defects}.
\newblock Applied Surface Science \textbf{285}(PARTB) (nov 2013)  858--864

\bibitem{LeCun}
LeCun, Y., Boser, B., Denker, J.S., Henderson, D., Howard, R.E., Hubbard, W.,
  Jackel, L.D.:
\newblock {Backpropagation Applied to Handwritten Zip Code Recognition}.
\newblock Neural Computation \textbf{1}(4) (1989)  541--551

\bibitem{Masci2012}
Masci, J., Meier, U., Ciresan, D., Schmidhuber, J., Fricout, G.:
\newblock {Steel defect classification with Max-Pooling Convolutional Neural
  Networks}.
\newblock In: The 2012 International Joint Conference on Neural Networks
  (IJCNN), IEEE (jun 2012)  1--6

\bibitem{Soukup}
Soukup, D., Huber-M{\"{o}}rk, R.:
\newblock {Convolutional neural networks for steel surface defect detection
  from photometric stereo images}.
\newblock Advances in Visual Computing \textbf{8887} (2014)  668--677

\bibitem{Ke2016}
Ke, W., Huiqin, W., Yue, S., Li, M., Fengyan, Q.:
\newblock {Banknote Image Defect Recognition Method Based on Convolution Neural
  Network}.
\newblock International Journal of Security and Its Applications \textbf{10}(6)
  (2016)  269--280

\bibitem{Faghih-Roohi}
Faghih-Roohi, S., Hajizadeh, S., N{\'{u}}{\~{n}}ez, A., Babuska, R., Schutter,
  B.D.:
\newblock {Deep convolutional neural networks for detection of rail surface
  defects}.
\newblock In: 2016 International Joint Conference on Neural Networks (IJCNN).
  (2016)  2584--2589

\bibitem{Park2016}
Park, J.K., Kwon, B.K., Park, J.H., Kang, D.J.:
\newblock {Machine learning-based imaging system for surface defect
  inspection}.
\newblock International Journal of Precision Engineering and
  Manufacturing-Green Technology \textbf{3}(3) (jul 2016)  303--310

\bibitem{Weimer2016}
Weimer, D., Scholz-Reiter, B., Shpitalni, M.:
\newblock {Design of deep convolutional neural network architectures for
  automated feature extraction in industrial inspection}.
\newblock CIRP Annals - Manufacturing Technology \textbf{65}(1) (jan 2016)
  417--420

\bibitem{Russakovsky2015}
Russakovsky, O., Deng, J., Su, H., Krause, J., Satheesh, S., Ma, S., Huang, Z.,
  Karpathy, A., Khosla, A., Bernstein, M., Berg, A.C., Fei-Fei, L.:
\newblock {ImageNet Large Scale Visual Recognition Challenge}.
\newblock International Journal of Computer Vision \textbf{115}(3) (2015)
  211--252

\bibitem{Kim2017}
Kim, S., Kim, W., Noh, Y.K., Park, F.C.:
\newblock {Transfer learning for automated optical inspection}.
\newblock In: 2017 International Joint Conference on Neural Networks (IJCNN),
  IEEE (may 2017)  2517--2524

\bibitem{Zhou2017}
Zhou, S., Chen, Y., Zhang, D., Xie, J., Zhou, Y.:
\newblock {Classification of surface defects on steel sheet using convolutional
  neural networks}.
\newblock Materiali in Tehnologije \textbf{51}(1) (2017)  123--131

\bibitem{goodfellow_explaining_2014}
Goodfellow, I.J., Shlens, J., Szegedy, C.:
\newblock {Explaining and Harnessing Adversarial Examples}.
\newblock Iclr 2015 (2015)  1--11

\bibitem{goodfellow_generative_2014-1}
Goodfellow, I., Pouget-Abadie, J., Mirza, M.:
\newblock {Generative Adversarial Networks}.
\newblock arXiv preprint arXiv: {\ldots} (2014)  1--9

\bibitem{Isola}
Isola, P., Zhu, J.Y., Zhou, T., Efros, A.A.:
\newblock {Image-to-Image Translation with Conditional Adversarial Networks}.
\newblock arXiv (2016) ~16

\bibitem{Zhu}
Zhu, J.Y., Park, T., Isola, P., Efros, A.A.:
\newblock {Unpaired image-to-image translation using cycle-consistent
  adversarial networks}.
\newblock arXiv preprint arXiv:1703.10593 (2017)

\bibitem{Shrivastava}
Shrivastava, A., Pfister, T., Tuzel, O., Susskind, J., Wang, W., Webb, R.:
\newblock {Learning from Simulated and Unsupervised Images through Adversarial
  Training}.
\newblock arXiv:112.07828 (2016) ~16

\bibitem{Canziani2016}
Canziani, A., Paszke, A., Culurciello, E.:
\newblock {An Analysis of Deep Neural Network Models for Practical
  Applications}.
\newblock (may 2016)

\bibitem{Ioffe2015}
Ioffe, S., Szegedy, C.:
\newblock {Batch Normalization: Accelerating Deep Network Training by Reducing
  Internal Covariate Shift}.
\newblock (feb 2015)

\bibitem{Hinton}
Hinton, G.E., Srivastava, N., Krizhevsky, A., Sutskever, I., Salakhutdinov,
  R.R.:
\newblock {Improving neural networks by preventing co-adaptation of feature
  detectors}.
\newblock (2012)

\bibitem{PyTorch}
PyTorch:
\newblock {PyTorch} (2017) [Online; accessed 29-November-2017].

\bibitem{Howard2013}
Howard, A.G.:
\newblock {Some Improvements on Deep Convolutional Neural Network Based Image
  Classification}.
\newblock arXiv preprint arXiv:1312.5402 (dec 2013)  1--6

\bibitem{Sokolova2009}
Sokolova, M., Lapalme, G.:
\newblock {A systematic analysis of performance measures for classification
  tasks}.
\newblock Information Processing and Management \textbf{45}(4) (2009)  427--437

\bibitem{Huang2016}
Huang, G., Liu, Z., Weinberger, K.Q., van~der Maaten, L.:
\newblock {Densely Connected Convolutional Networks}.
\newblock (aug 2016)

\end{thebibliography}
\end{document}